\documentclass[preprint,showpacs,preprintnumbers,amsmath,amssymb]{revtex4}

\usepackage{graphicx}
\usepackage{dcolumn}
\usepackage{bm}

\begin{document}

\preprint{}

\title{Fermi Liquid Model of Radiation Induced Magnetoresistance Oscillations in GaAs/Al$_x$Ga$_{1-x}$As Heterostructure Two-Dimensional Electron System: Theoretical Evidence of an Electron Reservoir 
}

\author{Tadashi Toyoda}
\thanks{email: toyoda@keyaki.cc.u-tokai.ac.jp}
\affiliation{
Department of Physics, Tokai University, Kitakaname 1117, Hiratsuka, Kanagawa, Japan 259-1292
}
\date{\today}

\begin{abstract}
The magnetoresistance oscillations in GaAs/Al$_x$Ga$_{1-x}$As heterostructures induced by millimeterwave radiation recently observed by Zudov et al. [Phys. Rev. B {\bf 64}, 201311 (2001)] is theoretically reproduced by introducing a model based on the finite temperature Fermi liquid theory of dc conductivity and the electron reservoir hypothesis. The obtained oscillation patterns show excellent agreement with the experimental result by Zudov et al. and are independent of the polarization of the radiation field in accordance with the experimental observation by Smet et al. [Phys. Rev. Lett. {\bf 95}, 116804 (2005)].
\end{abstract}



\maketitle

The new class of magnetoresistance oscillations in a high-mobility two-dimensional electron system (2DES) in a GaAs/Al$_x$Ga$_{1-x}$As heterostructure subjected to weak magnetic fields and millimeterwave radiation found by Zudov et al. \cite{Zudov2001}\cite{Zudov2003}\cite{Zudov2004} and by Mani et al. \cite{Mani2002}\cite{Mani2004} has revealed again the fascinating nature of the 2DES \cite{{Studenikin2005},{Dorozhkin2005},{Simovic2005},{Studenikin2007},{Wirthmann2007},{Hatke2008}}, evoking the discoveries of the quantum Hall effects (QHE) by von Klitzing \cite{vonKlitzing1980} and by Tsui et al. \cite{Tsui1982}. The period of these new oscillations is governed by the ratio of the millimeterwave to cyclotron frequencies, and the minima of the oscillations are characterized by an exponentially vanishing diagonal resistance. When the millimeterwave radiation is turned off, the magnetoresistance shows the well-established SdHvA oscillation whose period is governed by the ratio of the chemical potential to the cyclotron frequency. Hence this magnetoresistance oscillation is apparently induced by the illumination with millimeterwaves. Although various different theoretical models have been proposed \cite{{Ryzhy1969}, {Ryzhii1970}, {Ryzhii2003a}, {Ryzhii2003b}, {Ryzhii2003c}, {Ryzhii2004}, {Durst2003}, {Sedrakyan2008}, {Lei2004}, {Mikhailov2004}, {Andreev2003}, {Dietel2005}, {Torres2005}, {Robinson2004}, {Inarrea}, {Inarrea2007}, {Vavilov2004a}, {Vavilov2004b}, {Dmitriev}, {Dmitriev2005}, {Dmitriev2008}}, one crucial question remains to be solved. The experiment by Smet et al. \cite{Smet2005} shows that the resistance oscillations are notably immune to the polarization of the radiation field. This observation is discrepant with these theories and seems to cast doubt on the validity of the theoretical models so far proposed \cite{Zhang}. In these circumstances it seems necessary to explore other theoretical possibilities to explain the phenomena.

Meanwhile, an extremely interesting observation was made by Holland et al., who measured the long-wavelength magnetoplasmon dispersion in a high mobility 2DES realized in a GaAs/Al$_x$Ga$_{1-x}$As heterostructure quantum well \cite{Holland2004}. Using the coupling between the plasmon and THz radiation, they explored a wide range of filling factors for a fixed value of the wave-number vector to obtain an explicit filling-factor dependence of the dispersion. The observed dispersion shows a clear filling-factor dependent plateau-type dispersion, violently deviating from the well-established semiclassical formula at first glance. Their discovery suggests a previously unknown relation between the magnetoplasmon and the integer QHE (IQHE).
The full theoretical explanation of the phenomenon was given very recently \cite{Toyoda2008} by adopting the electron reservoir hypothesis (ERH) that was introduced more than two decades ago to explain the IQHE \cite{{Baraff1981},{Toyoda1984},{Toyoda1985}}.
That the experimental discovery by Holland et al. can be explained by adopting the ERH strongly indicates the significance of taking into account the electron reservoir in order to investigate the behavior of the electrons in the GaAs/Al$_x$Ga$_{1-x}$As heterostructure 2DES. In this paper a Fermi liquid model based on the ERH is proposed to explain the new class of the magnetoresistance oscillations within the framework of the standard quantum theory of electrical conductivity. The derived formula gives the magnetoresistance oscillation patterns which show excellent agreement with the experimental data observed by Zudov et al., including the SdHvA oscillation part.

The Fermi liquid theory of electrical conductivity was originally formulated by Eliashberg. The core of the theory is the analytic continuation of the finite temperature current correlation function with respect to the Matsubara frequency to obtain the retarded real-time current response function, which directly yields the conductivity. Here the formulation given in Ref.\cite{Toyoda1989} is applied to an electron gas confined in the $xy$-plane and subjected to a magnetic field ${\bf B}=(0,0,B)$ and a millimeter wave whose angular frequency is $\nu$. The polarization of the radiation field is arbitrary. The electron charge and effective mass are denoted by $-e$ and $m$, respectively. The interactions between the electrons and impurities or defects are treated perturbatively by the Bethe-Salpeter (BS) equation for the vertex function. The quasi-electron spectrum is assumed to be $\varepsilon=\hbar\omega_c(M+1/2)+(g^\ast \mu_B B/2){\rm sgn}\alpha \equiv \varepsilon_{M\alpha}$, where $M$ is the principal quantum number for the Landau levels, $\omega_c = eB/mc$ is the cyclotron frequency, $g^\ast$ is the effective $g$-factor, $\mu_B=eh/4\pi m_0 c$ is the Bohr magneton with the electron rest mass $m_0$, and the function sgn gives the sign of the spin variable $\alpha$. The Landau levels are degenerate and there is an additional quantum number $p$, which corresponds to the $x$-component of the electron momentum. The allowed range of this momentum is $\vert p \vert < p_{max} \equiv \pi e B/hc$. 

Although a microscopic model for the scattering mechanism responsible for the conductivity may be introduced via the proper vertex function in the BS equation, here we do not use a specific model but simply assume that the scattering mechanism is independent of $M$ and $\alpha$. Then applying the theory formulated in Ref.\cite{Toyoda1989} to the two-dimensional quasi-electrons, the general expression of conductivity can be found as
\begin{eqnarray}
\sigma_{xx} = \frac{-e^2 \hbar}{4m^2}\sum_\alpha \sum_{M=0}^\infty 
\frac{1}{2\pi} \int d\omega \int_{-p_{max}}^{p_{max}}dp \; p^2 
\frac{\partial f(\omega)}{\partial \omega} 
G^R_{M\alpha} (p,\omega) G^A_{M\alpha} (p,\omega) 
\nonumber \\ 
\times 
\left\{
1+\frac{1}{\hbar}{\rm Re}\Lambda_{\rm II}(p,\omega)
\right\},
\end{eqnarray}
where $f(\omega)= (1+\exp (\beta\hbar\omega))^{-1}$, $G^R$ ($G^A$) is the retarded (advanced) Green function, and 
$\Lambda_{\rm II}$ is the vertex function. The aim of this paper is to examine the general consequence of this conductivity formula when the ERH is adopted. 

The fundamental assumption of the Fermi liquid model, i.e., the Fermi liquid hypothesis (FLH), can be mathematically formulated in terms of the retarded and advanced Green functions, $G^R$ and $G^A$. In Ref.\cite{Toyoda1989}  it has been shown that by virtue of the FLH the product $G^R G^A$ in the above conductivity formula can be written as
\begin{eqnarray}
G^R (\omega) G^A (\omega) = 
\frac{\pi a^2}{\gamma} \delta
\left(
\omega - \hbar^{-1}(\varepsilon-\mu)
\right) 
+ \{ \textrm{non-singular term} \}
\;,
\end{eqnarray}
where $a$ is the wave function renormalization constant, and $\gamma$ is the damping parameter, which should satisfy the condition $\beta\hbar\gamma \ll 1$, and $\mu$ is the chemical potential. 

In the theories so far proposed \cite{{Ryzhy1969}, {Ryzhii1970}, {Ryzhii2003a}, {Ryzhii2003b}, {Ryzhii2003c}, {Ryzhii2004}, {Durst2003}, {Sedrakyan2008}, {Lei2004}, {Mikhailov2004}, {Andreev2003}, {Dietel2005}, {Torres2005}, {Robinson2004}, {Inarrea}, {Inarrea2007}, {Vavilov2004a}, {Vavilov2004b}, {Dmitriev}, {Dmitriev2005}, {Dmitriev2008}}, the effects of the millimeterwave radiation only on the 2DES were considered. However, if there is an electron reservoir, the effects of the radiation on the electrons in the reservoir should also be taken into account. At present the mechanism of the electron reservoir is not known. Here electrons in the reservoir are assumed to be in bound states with binding energy $-E_{res}$. Since the amount of the energy that an excited electron can receive from millimeterwave radiation is $\hbar\nu$, the condition that the electron can join the 2DES should be $\hbar\omega_c/2 + E_{res}<\hbar\nu$, which can also be written as $B<(2mc/\hbar e)(\hbar\nu - E_{res}) \equiv B_c$. As the chemical potential is the minimum free energy to add an electron to the system, the emergence of such a process may be described by introducing another singularity in the retarded Green function at $\hbar\nu$. This singularity of the Green function may be expressed as an effective chemical potential in the expression such as (2). These considerations lead to the following expression for $G^R G^A$: 
\begin{eqnarray}
G^R (\omega) G^A (\omega) =  
\sum_{i=1,2}
\frac{\pi a_i^2}{\gamma_i} 
\delta
\left( 
\omega-\hbar^{-1}(\varepsilon_{M\alpha}-\eta_i)
\right) 
\theta_i 
+\{ \textrm{non-singular term} \}
\end{eqnarray}
when the radiation is on, and 
\begin{eqnarray}
G^R (\omega) G^A (\omega) =  
\frac{\pi a_2^2}{\gamma_2} 
\delta
\left( 
\omega-\hbar^{-1}(\varepsilon_{M\alpha}-\eta_2)
\right) 
+\{ \textrm{non-singular term} \}
\end{eqnarray}
when the radiation is off. Here we have defined $\eta_1=\hbar\nu$, $\eta_2=\mu$, and regularized step functions $\theta_1 \equiv (1+\exp \tau (B - B_c ))^{-1}$ and $\theta_2 \equiv (1+\exp \tau (B_c -  B ))^{-1}$. The limit $\tau \rightarrow \infty$ corresponds to the sharp cut-off. By virtue of the delta functions the integration over $\omega $ in (1) for the two terms in (3) becomes trivial. Since the damping takes place for non-zero value of the momentum, the main contribution to the $p$-integration comes from the vicinity of $p=0$. That is, we can safely approximate $\gamma_i \simeq \gamma_{0i} p^2$. Then the conductivity formula (1) yields 
\begin{eqnarray}
\sigma_{xx} = \frac{-e^2\hbar}{4\pi m^2} 
\sum_\alpha 
\frac{eB}{ch} \sum_{M=0} 
\left\{
\lambda_1 f_\omega (\varepsilon_{M\alpha} - \hbar\nu)
\theta_1 + 
\lambda_2 f_\omega (\varepsilon_{M\alpha} - \mu)
\theta_2 + 
\lambda_0
\right\} 
\end{eqnarray} 
when the radiation is on, and
\begin{eqnarray}
\sigma_{xx} = \frac{-e^2\hbar}{4\pi m^2} 
\sum_\alpha 
\frac{eB}{ch} \sum_{M=0} 
\left\{
\lambda_2 f_\omega (\varepsilon_{M\alpha} - \mu)
+ \lambda_0
\right\} 
\end{eqnarray} 
when the radiation is off. 
The quantities 
${
\lambda_i \equiv a_i^2(8\pi \gamma_{0i})^{-1} 
\left\{
1+\hbar^{-1}{\rm Re} \Lambda_{\rm II}  
\right\} 
}$ 
for $i$=1 and 2 depend on the microscopic mechanism of the scattering process responsible for the conductivity. The contribution from the non-singular term is denoted by $\lambda_0$. For simplicity, we assume $\lambda_1 = \lambda_2 \equiv \lambda$. Using the explicit form of $f(\omega)$, the conductivity can be expressed as 
\begin{eqnarray}
\sigma_{xx}=\frac{e^2 \lambda}{m}
\left\{
W_1\theta_1 + W_2 \theta_2 + \xi B
\right\} \equiv \frac{e^2 \lambda}{m} W_{\rm ON} 
\end{eqnarray}
when the radiation is on, and 
\begin{eqnarray}
\sigma_{xx}=\frac{e^2 \lambda}{m}
\left\{
W_2 + \xi B
\right\} \equiv \frac{e^2 \lambda}{m} W_{\rm OFF} 
\end{eqnarray}
when the radiation is off. Here 
$\xi \equiv (-e/4 \pi^2 mc \lambda)\sum_M \lambda_0 $ 
is assumed to be a constant, and $W_i$'s are given as  
\begin{eqnarray}
W_i = 
\sum_\alpha \sum_{M=0}^\infty 
\frac{\beta \hbar \omega_c}
{
\left\{
1+e^{\beta(\varepsilon_{M \alpha}-\eta_i)}
\right\} 
\left\{
1+e^{-\beta(\varepsilon_{M \alpha}-\eta_i)}
\right\}
} \; .
\end{eqnarray}
This conductivity formula is derived within the theoretical framework of the linear response approximation, in which the electric field ${\bf E}=(E_x, E_y)$ is an externally controlled small perturbation. In the measurement by Zudov et al. \cite{Zudov2001} the current $I_x$ is measured by controlling the electric field $E_x$, while the current $I_y$ as well as the external electric field $E_y$ are kept zero. Therefore, the resistivity $R_{xx}$ observed in their measurement should correspond to $1/{\sigma_{xx}}$ in this theory. The significance of considering the boundary conditions on the current expectation values has been discussed in Ref.\cite{Toyoda1985}, where the IQHE formula is derived from a microscopic many-body Hamiltonian without depending on the linear response approximation. Thus, within the theoretical framework of the present formulation the resistivity corresponding to $R_{xx}$ in Ref.\cite{Zudov2001} is given as
\begin{eqnarray}
R_{xx} =
\sigma_{xx}^{-1} = \frac{m}{e^2 \lambda}
W^{-1}_{\rm{ON \; or \; OFF}}(B,T,\mu, \nu) \;.
\end{eqnarray}
The function $W_{\rm {ON}}$ depends on three adjustable parameters $B_c$, $\tau$, and $\xi$, while $W_{\rm {OFF}}$ depends only on  $\xi$. These parameters will be adjusted to optimize the fit of the theoretical curve with the experimental data. 

The chemical potential of 2DES without millimeterwaves is given as 
\cite{Toyoda2008}
\begin{eqnarray}
\mu =\frac{\pi \hbar^2 n}{m}\;,
\end{eqnarray}
where $n$ is the electron number density in the limit of very weak magnetic field. This gives a good approximation for the range of the parameters in the experiments of Ref.\cite{Zudov2001}. It should be noted here that the above approximation gives the classical Hall effect instead of the IQHE. This explains the experimental observation reported in Ref.\cite{Zudov2004}. 

The $B$-dependence of $W^{-1}_{\rm ON}$ for $0.2 < B < 2$ kG is shown by a solid line in Fig. 1, using the values of the physical parameters given in Ref.\cite{Zudov2001}, i.e., $T = 0.5$ K, $n = 2 \times 10^{11}$cm$^{-2}$, $m=0.068m_0$, and $f = 100$ GHz. The adjustable parameters are chosen such that $B_c =  3.0 $ kG, $\tau = 0.01$, and $\xi = 1.8 \times 10^{-4} $. These values indicate $E_{res}\simeq 0.16$(meV). The experimentally measured $R_{xx}$ taken from Fig. 4 in Ref.\cite{Zudov2001} is also shown by a dotted line for the same $B$ values.  The oscillation pattern of $W_{\rm ON}^{-1}$ shows excellent agreement with the experimental result. In Fig. 2 the $B$-dependence of  $W^{-1}_{\rm ON}$  is shown by a solid line for $0.2 < B < 3.6$ kG, using the same values for the parameters. The experimentally measured $R_{xx}$ taken from Fig. 4 in Ref.\cite{Zudov2001} is also shown by a dotted line for the same $B$ values. The theoretical pattern shows excellent agreement with the expermental curve.
The phase of the oscillation observed in the experimental curve is shifted toward positive $B$ direction. The shift becomes prominent for $B > 2.2$ kG. Theoretically, an additional term to the quasi-electron energy spectrum, i.e., the proper self-energy, may cause such a shift. The magnetoresistivity data given in Fig. 1 of Ref.\cite{Zudov2001} show a peak due to the magnetoplasmon resonance around $B \sim 2.4$ kG. The existence of magnetoplasmon resonance peak in the magnetoresistivity in Hall bar structures was first observed by Vasiliadou et al. \cite{Vasiliadou1993} in the microwave photoconductivity measurement. The magnetoplasmon resonance gives considerable contribution to the two-particle Green function, which is directly related to the proper self-energy of the single particle excitation spectrum via the finite temperature Ward-Takahashi relation \cite{Toyoda1989}. In Fig. 3, the function $W_{\rm OFF}^{-1}$ is plotted for $0.2 < B < 3.6$ GHz. It exhibits clear SdHvA oscillation observed experimentally.

In conclusion, the $B$-dependence of the oscillatory patterns of the millimeterwave induced magnetoresistance oscillations observed by Zudov et al. is almost perfectly reproduced from our theoretical model based on the FLH and the ERH. Furthermore, it should be noted that the present model is independent of the polarization of the raditation field in perfect accordance with the experimental observation by Smet et al. \cite{Smet2005}. Since the model does not depend on any specific mechanisms of the electron scatterings that cause electric resistivity, this new magnetoresistivity oscillation seems to be a universal consequence of the Fermi liquid nature of the 2DES and the existence of an electron reservoir in GaAs/AlGaAs heterostructures. 

In view of the fact that the present theory can reproduce excellently the experimental data including the immunity to the polarization of the radiation field, the next task should be to reveal the microscopic mechanism that realizes the electron reservoir. At present there seem to be no direct experimental clues to that question. Nevertheless, it should be noted that there has been no direct experimental evidence against the existence of an electron reservoir either. In other words there has been no direct experimental evidence that the electron number is fixed. 

As the essential theoretical conjecture of this work, i.e., Eqs. (3) and (4), can be used to calculate various physically measurable quantities, the validity of the model presented in this work may be further tested.

\begin{acknowledgments}
I thank Chao Zhang for pointing out the significance of this theory on the immunity of the radiation induced magnetoresistance oscillation to the polarization of the radiation field and bringing Ref. \cite {Smet2005} to my attention. I also thank Masaki Yasu\`{e} and Wolfgang Bentz for valuable comments.
\end{acknowledgments}

\newpage
\noindent
{\bf Fgiure captions}\\
\\
Fig. 1\\
The $B$-dependence of the function $W^{-1}_{\rm ON}$  and the experimentally measured $R_{xx}$ for $0.1<B<2$ when the system is illuminated with radiation. The parameters are: $f=$100 GHz, $T = 0.5$ K, $n = 2 \times 10^{11}$cm$^{-2}$ . (a) Theoretical $W^{-1}_{\rm ON}$ defined by Eq. (7) is plotted by the solid line. (b) Experimentally measured $R_{xx}$ taken from Fig. 4 in Ref. \cite{Zudov2001} is plotted by the dotted line. 
\\
\\
Fig. 2\\
The $B$-dependence of the function $W^{-1}_{\rm ON}$ and the experimentally measured $R_{xx}$ for $0.2<B<3.6$ when the system is illuminated with radiation. The parameters are same as Fig. 1. (a) Theoretical $W^{-1}_{\rm ON}$ defined by Eq. (7) is plotted by the solid line.  (b) Experimentally measured $R_{xx}$ taken from Fig. 4 in Ref. \cite{Zudov2001} is plotted by the dotted line.  
\\
\\
Fig. 3\\ 
The $B$-dependence of the function $W^{-1}_{\rm OFF}$ for $0.2<B<3.6$ when the system is not illuminated with radiation. It shows a typical SdHvA oscillation.

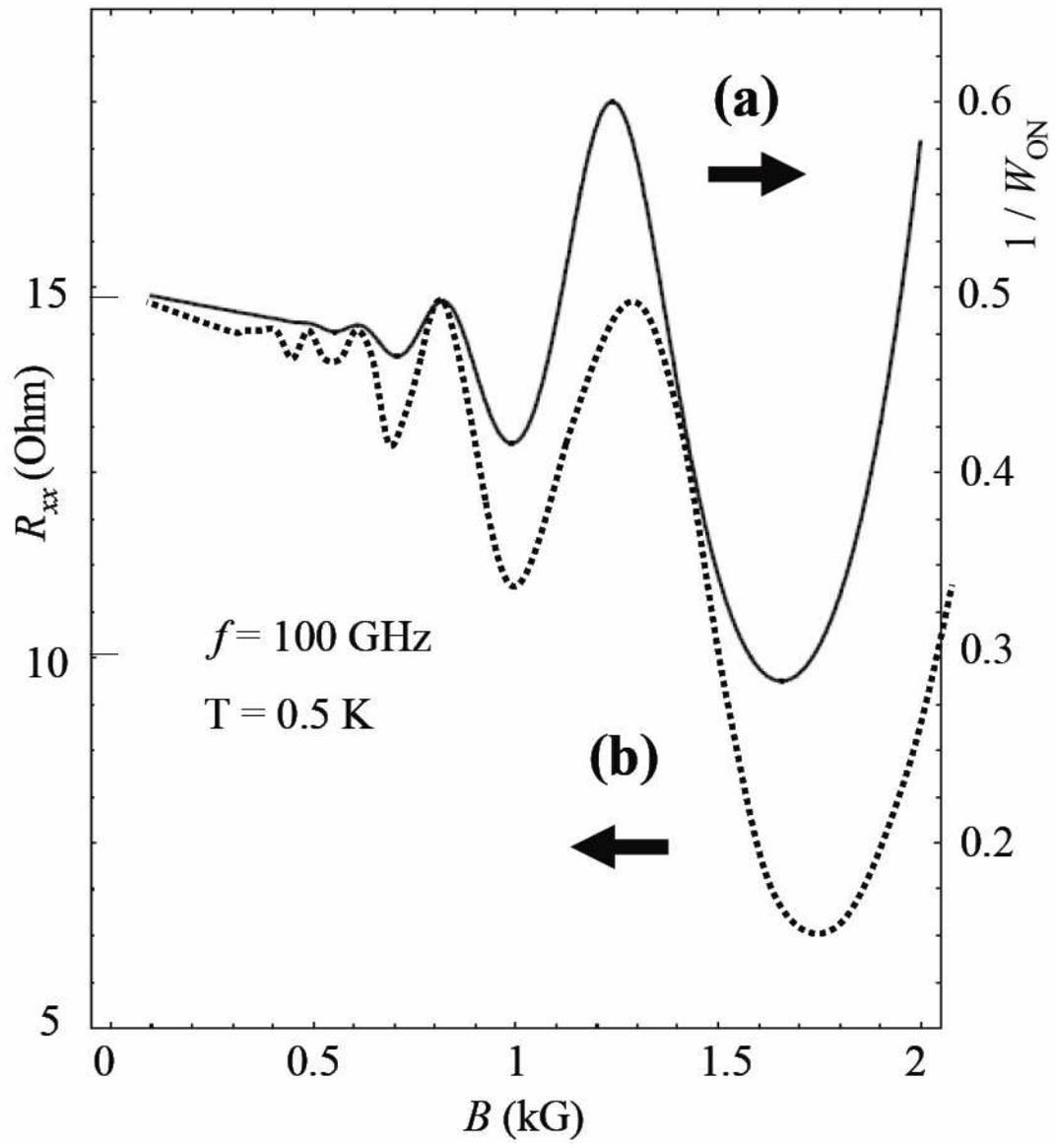
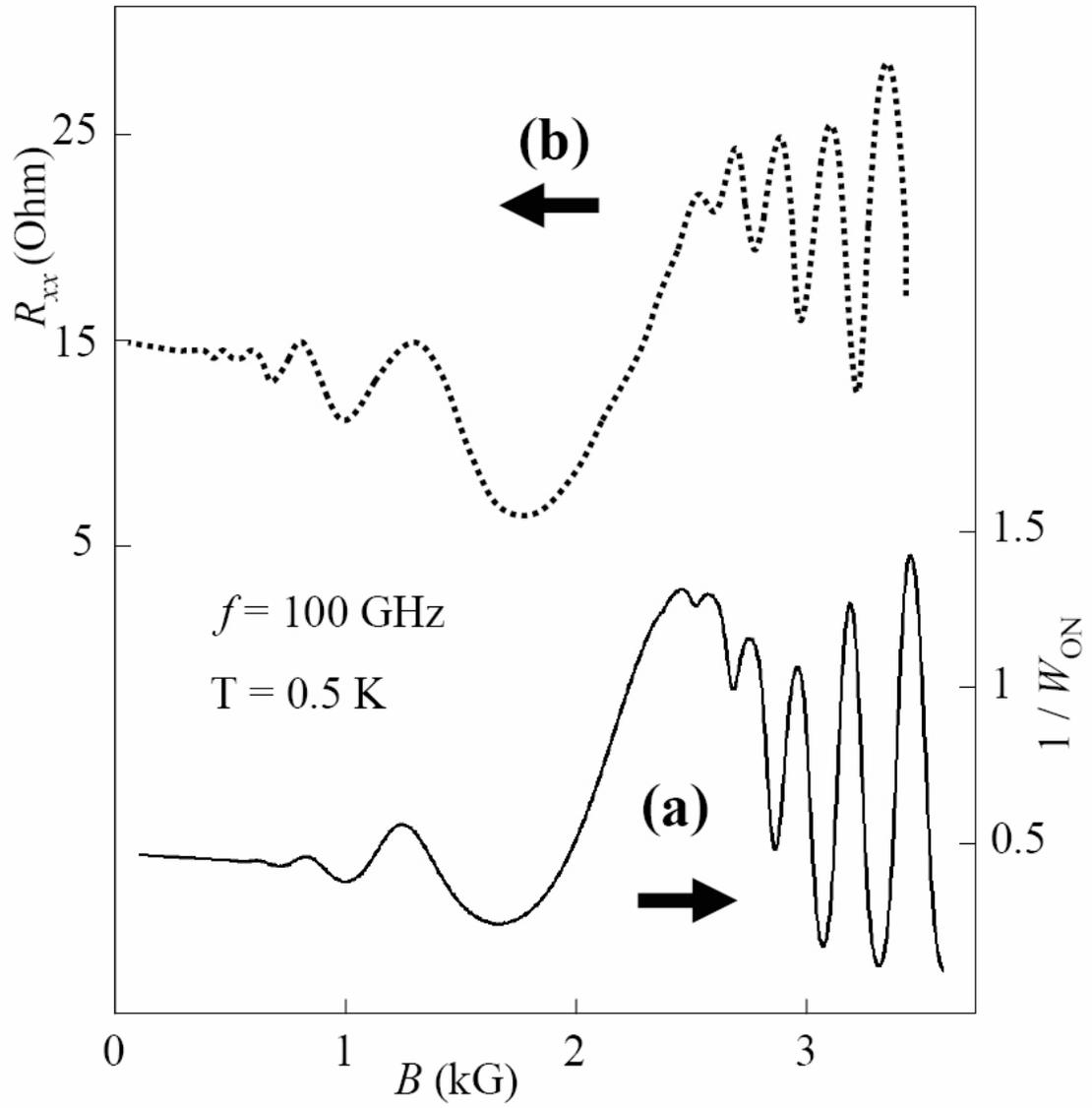
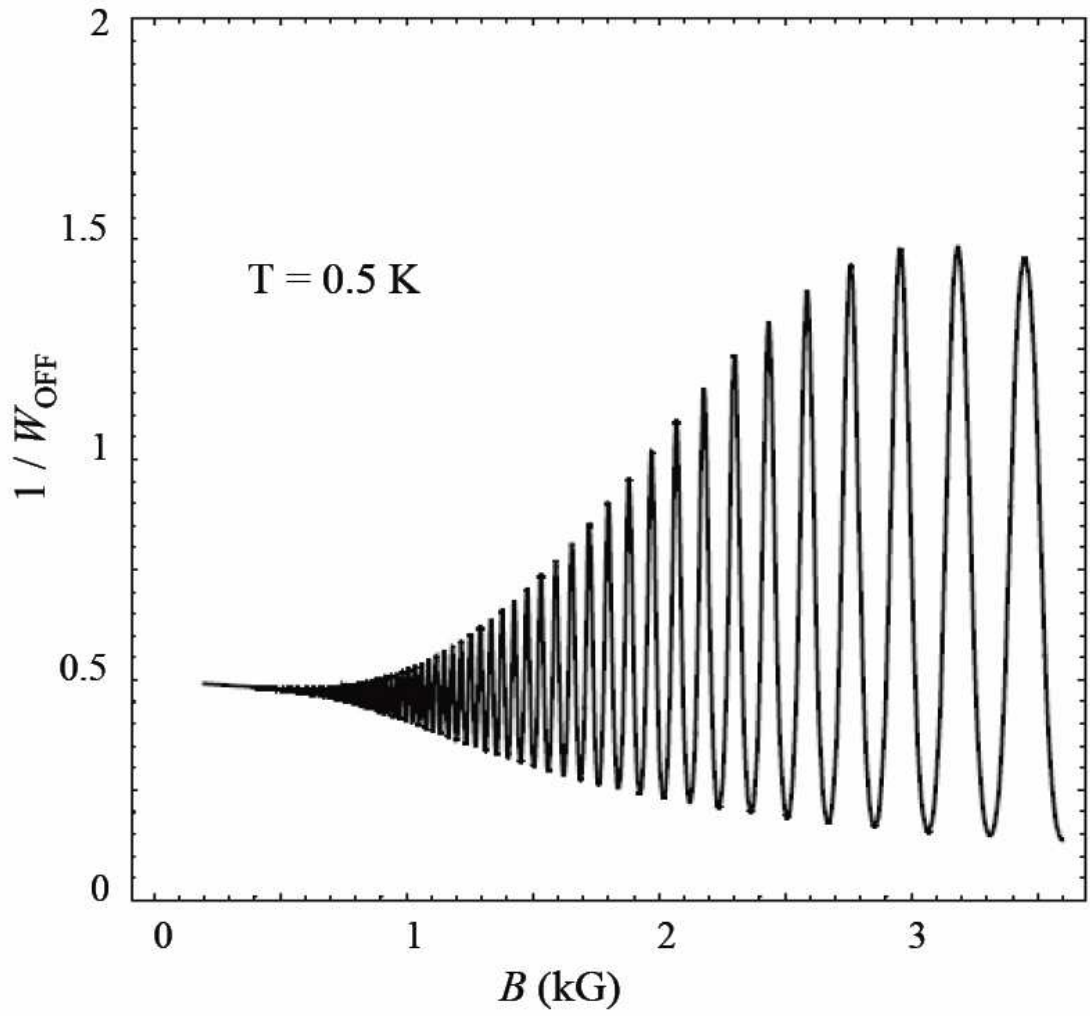
\end{document}